\documentclass[default,iicol,sn-mathphys,Numbered]{sn-jnl}

\usepackage{graphicx}%
\usepackage{multirow}%
\usepackage{amsmath,amssymb,amsfonts}%
\usepackage{amsthm}%
\usepackage{mathrsfs}%
\usepackage[title]{appendix}%
\usepackage{xcolor}%
\usepackage{textcomp}%
\usepackage{manyfoot}%
\usepackage{booktabs}%
\usepackage{algorithm}%
\usepackage{algorithmicx}%
\usepackage{algpseudocode}%
\usepackage{listings}%
\usepackage{adjustbox}
\usepackage[normalem]{ulem}

\def\Jpsi   {J\mbox{\footnotesize$\!/\psi$}}

\theoremstyle{thmstyleone}%
\theoremstyle{thmstyletwo}%

\theoremstyle{thmstylethree}%

\begin{document}

\title[Review of Semileptonic $B$ Anomalies]{Review of Semileptonic $B$ Anomalies}

\author*[1,2]{\fnm{Bernat} \sur{Capdevila}}\email{bcapdevila@ifae.es}

\author*[3,4]{\fnm{Andreas} \sur{Crivellin}}\email{andreas.crivellin@psi.ch}

\author*[2]{\fnm{Joaquim} \sur{Matias}}\email{matias@ifae.es}

\affil[1]{\orgdiv{DAMTP}, \orgname{University of Cambridge}, \orgaddress{\street{Wilberforce Road}, \city{Cambridge}, \postcode{CB3 0WA}, \state{United Kingdom}}}

\affil[2]{\orgdiv{Department of Physics and IFAE}, \orgname{Universitat Aut\`onoma de Barcelona}, \orgaddress{\street{Edifici Ciències}, \city{Bellaterra}, \postcode{08193}, \state{Barcelona}, \country{Spain}}}

\affil[3]{\orgdiv{Physik-Institut}, \orgname{Universit\"at Z\"urich}, \orgaddress{\street{Winterthurerstrasse 190}, \city{Z\"urich}, \postcode{8057}, \state{Z\"urich}, \country{Switzerland}}}

\affil[4]{\orgdiv{LTP}, \orgname{Paul Scherrer Institut}, \orgaddress{\street{Foruschungsstr.~101}, \city{Villigen}, \postcode{5232}, \state{Aargau}, \country{Switzerland}}}

\abstract{\unboldmath We review the current status and implications of the anomalies (i.e. deviations from the Standard Model predictions) in semi-leptonic $B$ meson decays, both in the charged and in the neutral current. In $b\to s\ell^+\ell^-$ transitions significant tensions between measurements and the Standard Model predictions exist. They are most pronounced in the branching ratios ${\cal B}_{B \to K\mu^+\mu^-}$ and ${\cal B}_{B_s\to\phi\mu^+\mu^-}$ (albeit quite dependent on the form factors used) as well as in angular observables in $B\to K^*\mu^+\mu^-$ (the $P_5^\prime$ anomaly). Because the measurements of ${\cal B}_{B_s\to \mu^+\mu^-}$ and of the ratios $R_K$ and $R_{K^*}$ agree reasonably well with the SM predictions, this points towards (dominantly) lepton flavour universal New Physics coupling vectorially to leptons, i.e. contributions to $C_9^{\rm U}$. In fact, global fits prefer this scenario over the SM hypothesis by $5.8\sigma$. Concerning $b\to c\tau\nu$ transitions, $R(D)$ and $R(D^*)$ suggest constructive New Physics at the level of $10\%$ (w.r.t. the Standard Model amplitude) with a significance above $3\sigma$. We discuss New Physics explanations of both anomalies separately as well as possible combined explanations. In particular, a left-handed vector current solution to $R(D^{(*)})$, either via the $U_1$ leptoquark or the combination of the scalar leptoquarks $S_1$ and $S_3$, leads to an effect in $C_9^{\rm U}$ via an off-shell penguin with the right sign and magnitude and a combined significance (including a tree-level effect resulting in $C_{9\mu}^\mathrm{V}=-C_{10\mu}^\mathrm{V}$ and $R(D^{(*)})$) of $6.3\sigma$. Such a scenario can be tested with $b \to s \tau^+\tau^-$ decays. Finally, we point out an interesting possible correlation of $R(D^{(*)})$ with non-leptonic $B$ anomalies.}


\maketitle	
\section{Introduction} 

The Cabibbo-Kobayashi-Maskawa (CKM) mechanism~\cite{Kobayashi:1973fv}, encoded in the Standard Model (SM) of particle physics, was established by the $B$ factories BELLE~\cite{Belle:2000cnh} and BaBar~\cite{BaBar:2001yhh} in the first decade of this century to be the leading source of quark flavour violation. Furthermore, the discovery of the Brout-Englert-Higgs boson~\cite{Higgs:1964ia,Englert:1964et} in 2012 at the Large Hadron Collider (LHC) at CERN~\cite{ATLAS:2012yve,CMS:2012qbp} completed the Standard Model (SM) of particle physics. Therefore, the focus of current research shifted even more towards unveiling new particles and new interactions, not contained within the SM.

New particles can be searched for both directly at high energies and indirectly at the low-energy precision frontier. While the LHC experiments ATLAS and CMS only recently observed interesting direct hints for new particles~\cite{Fischer:2021sqw}, in particular for new scalar bosons at 95\,GeV and 151\,GeV~\cite{Bhattacharya:2023lmu}, historically indirect discoveries have often preceded direct ones. For example, the existence of the charm quark, the $W$ boson, the top quark and also the Higgs were previously indicated by indirect measurements, like Fermi interactions, Kaon mixing and electroweak precision observables. 

In this context, semi-leptonic $B$ meson decays are a particularly useful tool for indirect searches: they have in general clean experimental signatures, controllable theoretical uncertainties and suppressed rates, which makes them very sensitive probes of new physics (NP). In this report, we will focus on the most pronounced anomalies which are found in $b\to s \ell^+\ell^-$ and $b\to c\tau \nu$ processes. On the one side, $b\to c\tau \nu$ is a tree-level charged current measured dominantly in the ratios of branching ratios of ${\cal B}_{B\to D^{(*)}\tau\nu}$ and ${\cal B}_{B\to D^{(*)}\ell\nu}$ which point towards the violation of lepton flavour universality (LFU) satisfied by the SM gauge interactions. On the other side, the transition $b\to s \ell^+\ell^-$ is a flavour-changing neutral current process that is only generated in the SM at the loop level. Here the most relevant observables are in the decays $B\to K^{(*)} \mu^+\mu^-$, $B_s\to\mu^+\mu^-$ and $B_s \to \phi \mu^+\mu^-$, including optimized angular observables ($P_i$)~\cite{Descotes-Genon:2013vna} for $B\to K^*\mu^+\mu^-$ or $B_s \to \phi\mu^+\mu^-$. 

As NP can in the large majority of cases be assumed to be heavy compared to the $B$ meson scale (see e.g.~Refs.~\cite{Sala:2017ihs,Crivellin:2022obd} for some studies of light new particles in $b\to s\ell^+\ell^-$), the theoretical framework employed for such analyses is the Weak Effective Theory (WET)~\cite{Witten:1976kx,Buchalla:1995vs}. In this setup, the electroweak scale SM particles (top quark, $W$, $Z$ and the Higgs), as well as potential beyond the SM fields, are integrated out. This allows one to describe these transitions in terms of an effective Hamiltonian model and process-independent interactions.

In this review, we will discuss the experimental and theoretical status of these anomalies in Sec.~\ref{bsll} and Sec.~\ref{bctaunu} including the latest model-independent fits. We will then review possible NP explanations of the anomalies in Sec.~\ref{npexplanations}, first for the two classes separately, and afterwards consider combined explanations. We will conclude in Sec.~\ref{correlations} with an outlook on future developments and highlight complementary observables, like $b\to s\tau^+\tau^-$ processes and anomalies in non-leptonic $B$ decays with emphasis on some possible links with the semi-leptonic ones. 

\section{$b\to s\ell^+\ell^-$
}\label{bsll}

In the analysis of $b\to s\ell^+\ell^-$ transitions, several decay channels (mostly exclusive ones) and their associated observables are considered. Exclusive decays involving pseudo-scalar mesons in the final state, such as $B^{(+,0)}\to K^{(+,0)}\ell^+\ell^-$, are important in $b\to s\ell^+\ell^-$ analyses. Their corresponding observables include the branching ratios $\mathcal{B}_{B^{(+,0)}\to K^{(+,0)}\ell\ell}$ and angular observables, such as the forward-backward asymmetry $A_{FB}^\ell$ and the \textit{flat term} $F_H^\ell$ (see Ref.~\cite{Bobeth:2001sq,Bobeth:2007dw} for definitions). 

Other exclusive channels that play a crucial role are decays to vector mesons such as $B^{(+,0)}\to K^{*(+,0)}\ell^+\ell^-$ and $B_s\to \phi\ell^+\ell^-$. The observables in these channels do not only include total branching ratios $\mathcal{B}_{B^{(+,0)}\to \left\{K^{*(+,0)},\phi\right\}\ell^+\ell^-}$ but also e.g.~longitudinal polarization fractions ($F_{L}$), and angular distributions. Their angular distributions can be described using optimised observables $P_{1,2,3,4,5,6,8}^{(\prime)}$~\cite{Descotes-Genon:2013vna,
Aaij:2015yra},
which reduce the sensitivity to hadronic uncertainties~\cite{Kruger:2005ep,Matias:2012xw,Egede:2010zc}. This is achieved by using the soft-collinear-effective-theory relations that emerge in the large recoil limit of $B\to V\ell^+\ell^-$ decays~\cite{Beneke:2000wa,Beneke:2001at}, where the vector meson $V$ is maximally recoiling against the parent $B$ meson. In this limit, the form-factor (FF) dependence at leading order in ${\cal O}(\Lambda/m_b, \alpha_s)$ (corresponding to the so-called soft FFs) cancels in the optimised observables, resulting also in an enhanced sensitivity to short-distance effects~\cite{Matias:2012xw,Descotes-Genon:2013vna}.

Furthermore, ratios testing LFU~\cite{Bobeth:2007dw,Hiller:2003js} such as
\begin{equation}
R({K^{(*)}})=\dfrac{\mathcal{B}_{B\to K^{(*)}\mu^+\mu^-}}{\mathcal{B}_{B\to K^{(*)}e^+ e^-}}\,,
\end{equation}
can be constructed. Since the weak interactions and QED treat all three lepton generations equally, significant departures from LFU would unequivocally indicate the presence of NP.

Also exclusive radiative decays, such as $B\to \left\{K^{(*)},\phi\right\}\gamma$, and inclusive decays, such as $B\to X_s\gamma$ and $B\to X_s\ell^+\ell^-$, are important $b\to s\ell^+\ell^-$ transitions. While the inclusive decays offer complementary constraints on the NP structures, the radiative decays impose strong constraints on the electromagnetic and chromomagnetic dipole operators. Finally, the purely leptonic decay $B_s\to\mu^+\mu^-$ stands out as one of the most stringent constraints on axial, scalar, and pseudo-scalar NP contributions.

\subsection{Anomalies}

Among the different tensions found in the last ten years in the observables governed by the $b \to s \ell^+\ell^-$ transitions mentioned above, one of the most persistent tensions is the one associated with the optimised observable $P_5^\prime$ of the 4-body decay $B\to K^{*}(\to K\pi)\mu^+\mu^-$. This observable deserves a dedicated discussion given its importance. 

The angular observable $P_5^\prime$~\cite{Descotes-Genon:2012isb} was first measured  at LHCb~\cite{LHCb:2013ghj} and later on at Belle~\cite{Belle:2016xuo,Belle:2016fev}, ATLAS~\cite{ATLAS:2018gqc} and CMS~\cite{CMS:2017rzx}. The most recent SM prediction for this observable 
in the two most anomalous bins within the approach described in Ref.~\cite{Capdevila:2017ert}
is presented in Ref.\cite{Alguero:2023jeh}.
After corresponding updates with the most recent calculation of $B\to V$ FFs with $B$-meson light-cone distribution amplitudes (LCDAs) up to twist four for the two- and three-particle distributions (the so-called GKvD FFs~\cite{Gubernari:2018wyi})\footnote{In Ref.~\cite{Gao:2019lta}
the NLO corrections to the leading-power contributions together with the subleading-power effects at twist-6 accuracy were computed.}
, read~\cite{Alguero:2023jeh}
\begin{align}
&P_{5\;\text{SM}}^{\prime\;[4.0, 6.0]} =-0.72 \pm 0.08  \;\; \\
&P_{5\;\text{SM}}^{\prime\;[6.0, 8.0]} =-0.81 \pm 0.08  \;\; 
\end{align}
The theoretical uncertainty budget for $P_5^\prime$ can be split into a parametric uncertainty, soft FFs, factorizable power corrections, non-factorizable power corrections and a long-distance charm-loop uncertainty~\cite{Capdevila:2017ert}. For different sets of FFs that we have implemented, we observed that the parametric ($\sim 30\%$) and factorisable power corrections ($\sim 50\%$) are the main sources of uncertainty. See Ref.~\cite{Capdevila:2017ert} for detailed definitions of each type of uncertainty. 

Even if the use of the new GKvD FFs has led to a reduction of part of the uncertainties associated with the FFs compared to the previously used KMPW FFs~\cite{Khodjamirian:2010vf}, the dominance of the conservative factorisable power corrections used in Ref.~\cite{Capdevila:2017ert} and the small shift in the SM central value prediction towards experiment has slightly decreased the tension with data.  
Still, the deviations between theory and experiment in these bins remain noteworthy:
\begin{align} \label{p5pn1}
&P_{5\;\text{LHCb}}^{\prime\;[4.0, 6.0]} =-0.439 \pm 0.111 \pm 0.036 \;\; (1.9\sigma),\\ \label{p5pn2}
&P_{5\;\text{LHCb}}^{\prime\;[6.0, 8.0]} =-0.583 \pm 0.090 \pm 0.030 \;\; (1.9\sigma).
\end{align}
Moreover, it is important to emphasize that the specific treatment of the factorisable power corrections has an impact on the size of the uncertainty on the predictions of $P_5^\prime$. For instance, if one chooses to use GKvD form factors including all correlations among power corrections, that are very dependent on the details and assumptions used in the LCSR computation, the uncertainty is smaller and the tension larger: 2.4$\sigma$ and 2.1$\sigma$, respectively. Instead if one takes a more skeptical stance and removes all information on the correlations among the factorizable power corrections, effectively removing part of the model dependence of the LCSR computation, and enhances their nominal size according to their power counting of ${\cal O}(\Lambda/m_b)$, see Refs.~\cite{Descotes-Genon:2014uoa,Capdevila:2017ert} for specific details on this framework, then naturally the associated uncertainty becomes larger and the tension smaller, c.f. Eqs.~\ref{p5pn1} and \ref{p5pn2}. Notice also that, within this latter approach, the leading contributions to the different FFs, i.e. the ones that emerge in the $m_b, E_H\to\infty$ limit (being $E_H$ the energy of the daughter meson emitted in the decay), are correlated according to the robust symmetries dictated by the aforementioned limits.

Besides parametric and FFs uncertainties, long-distance charm-loop contributions are also an important source of uncertainty. The  amplitude for the decay $B\to M\ell^+\ell^-$ has the following structure in the SM: 
\begin{eqnarray}
A(B\to M\ell^+\ell^-)&=&\frac{G_F \alpha}{\sqrt{2}\pi} V_{tb}V_{ts}^* [({A_\mu}+{ T_\mu}) \bar{u}_\ell \gamma^\mu v_\ell \nonumber \cr
 && +{B_\mu} \bar{u}_\ell \gamma^\mu \gamma_5 v_\ell]\,,
\end{eqnarray}
where ${A_\mu} = -(2m_bq^\nu/{q^2}) {\cal C}_7 \langle M |\bar{s}\sigma_{\mu\nu}P_R b|B\rangle +{\cal C}_9 \langle M |\bar{s}\gamma_\mu P_L b|B\rangle $ and ${B_\mu} = {\cal C}_{10}\langle M | \bar{s}\gamma_\mu P_L b|B\rangle$ (see Sec.~\ref{globalfits} for definitions of the Wilson coefficients ${\cal C}_{7,9,10}$). While the local contributions coming from FFs, discussed above, are included in $A_\mu$ and $B_\mu$, the non-local charm-quark loop contributions enter the piece $T_\mu$. The difficulty to disentangle a NP contribution in ${\cal C}_9$ and ${\cal C}_7$ from a non-local contribution to $T_\mu$ comes precisely because of their common contribution to the leptonic vectorial current. There are different approaches to model this non-local contribution:
\begin{itemize}
\item[\textit{i})] Using  LCSR to compute the leading one soft-gluon exchange \cite{Khodjamirian:2010vf}, confirmed  later on in \cite{Gubernari:2020eft} finding a rather small NLO  correction.

\item[\textit{ii})] Via a dispersive representation using $J/\Psi$ and $\Psi(2S)$ data to determine the analytic structure and $q^2$ dependence of the $T_\mu$ term~\cite{Bobeth:2017vxj}. A more recent and detailed analysis can be found in Ref.~\cite{Gubernari:2023puw}.

\item[\textit{iii})] A fit to the resonances in modulo and phase to check if the tail of a resonance could explain the deviation in the anomalous bins~\cite{Blake:2017fyh}. 
\end{itemize}
All these different approaches arrive at the same conclusion, namely, that such non-local contribution cannot explain the deviation in $P_5^\prime$.

Finally, on the experimental side, one should include also the tensions observed at 
Belle~\cite{Belle:2016xuo,Belle:2016fev}, ATLAS~\cite{ATLAS:2018gqc} and CMS~\cite{CMS:2017rzx}
 in $P_5^\prime$ but also on the charged channel $B^+\to K^{*+}\ell^+\ell^-$\cite{LHCb:2020gog} where some smaller tensions at the level of 1.1$\sigma$ and 1.6$\sigma$ (see Ref.\cite{Alguero:2023jeh}) are observed further increasing the tension in the anomalous bins. Section~\ref{globalfits} is devoted precisely to exploring the implications of combining all data.

The branching ratios for the different $b\to s\ell^+\ell^-$ modes under discussion represent another set of observables that show systematic and coherent tensions~\cite{LHCb:2014cxe,LHCb:2016ykl,LHCb:2021zwz} w.r.t.~the deviations observed in the angular distributions. 

On the one hand, during these last years, the LHCb collaboration has performed new experimental measurements of the branching ratio for the $B_s \to \phi \mu^+ \mu^-$ mode~\cite{LHCb:2021zwz}. Despite the significant tensions reported at the time of the measurement, with up to 3.6$\sigma$ when computing the corresponding theoretical predictions using fits to BSZ LCSRs and lattice data for the form factors~\cite{LHCb:2021zwz,Bharucha:2015bzk}, the theoretical predictions obtained within the framework of Ref.~\cite{Alguero:2023jeh}, that employs GKvD FFs without including lattice input, lead to a reduction of the tension w.r.t.~its SM prediction:
%
%
%
\begin{align}
\begin{aligned}
{\mathcal{B}_{B_s \to \phi \mu^+ \mu^-}^{[1.1,6.0],\text{SM}}}= &\;(
5.25 \pm 2.76) \times 10^{-8},\\
{\mathcal{B}_{B_s \to \phi \mu^+ \mu^-}^{[1.1,6.0],\text{LHCb}}} = &\;(2.88 \pm 0.15 \pm 0.05\\
&\;\pm 0.14) \times 10^{-8} \;\; (0.9\sigma).
\end{aligned}
\end{align}
Instead if one uses the approach of Ref.~\cite{Gubernari:2022hxn} to compute the normalised branching ratio for this channel, where the $B_s\to \phi J/\psi$ decay is used as the normalisation, one finds significant tensions above 2$\sigma$~\cite{Gubernari:2022hxn} in some bins. Notice that this normalisation is also used in the experimental analyses to help cancelling systematics. The computation of the normalised $B_s\to \phi\mu^+\mu^-$ branching ratio has been possible due to the treatment of hadronic uncertainties developed in Refs.~\cite{Gubernari:2020eft,Gubernari:2022hxn}. 

On the other hand, while there have been no experimental updates in recent times on the branching ratios of the $B\to K\mu^+\mu^-$ and $B\to K^*\mu^+\mu^-$ modes, significant progress on the theory side has impacted the predictions of the observables for the pseudoscalar channels. In particular, a new lattice calculation of the $B\to K$ form factors across the entire $q^2$ range by the HPQCD collaboration~\cite{Parrott:2022rgu} has substantially increased the precision of our computations of $B^{(+,0)}\to K^{(+,0)}\mu^+\mu^-$ related observables. Consequently, the uncertainties on the branching ratios of these channels in the low-$q^2$ region have been reduced from approximately $\mathcal{O}(30\%)$ to $\mathcal{O}(10\%)$, with no significant shifts in the central values. This improvement has led to tensions of $O(4\sigma)$ in several $q^2$ bins \cite{Alguero:2023jeh}:
\begin{align}
{\mathcal{B}_{B^+ \to K^+\mu^+ \mu^-}^{[1.1,2.0],\text{SM}}}= &\;(
0.33 \pm 0.03) \!\times\! 10^{-7},\nonumber\\
{\mathcal{B}_{B^+ \to K^+\mu^+ \mu^-}^{[4.0,5.0],\text{SM}}}= &\;(
0.37 \pm 0.03) \!\times\! 10^{-7},\nonumber\\
{\mathcal{B}_{B^+ \to K^+\mu^+ \mu^-}^{[5.0,6.0],\text{SM}}}= &\;(
0.37 \pm 0.03) \!\times\! 10^{-7},\\
{\mathcal{B}_{B^+ \to K^+\mu^+ \mu^-}^{[1.1,2.0],\text{LHCb}}} = &\;(
0.21 \pm 0.02) \!\times\! 10^{-7} \;(4.0\sigma),\nonumber\\
{\mathcal{B}_{B^+ \to K^+\mu^+ \mu^-}^{[4.0,5.0],\text{LHCb}}}= &\;(
0.22 \pm 0.02) \!\times\! 10^{-7} \;(4.4\sigma),\nonumber\\
{\mathcal{B}_{B^+ \to K^+\mu^+ \mu^-}^{[5.0,6.0],\text{LHCb}}} = &\;(
0.23 \pm 0.02) \!\times\! 10^{-7} \;(4.0\sigma).\nonumber
\end{align}
Conversely, the SM predictions for the branching ratio of $B\to K^*\ell^+\ell^-$ modes in the low-$q^2$ bins have experienced a remarkable reduction in uncertainties compare Ref.~\cite{Alguero:2023jeh} to Ref.~\cite{Alguero:2021anc}, and now exhibit lower central values. These improvements arise from recent advancements in SM predictions  that rely on updated  GKvD $B\to K^*$ FFs, resulting in SM predictions for the branching ratios that are in fairly good agreement with experimental results.

Including the latest CMS measurement~\cite{CMS:2022mgd} the purely leptonic decay $B_s\to\mu^+\mu^-$  has a branching ratio of
\begin{equation}
\mathcal{B}^\text{Av.}
_{B_s\to\mu^+\mu^-} = (3.52^{+0.32}_{-0.30})\times 10^{-9} \;\; .
\end{equation}
This updated average places the branching ratio at around $1\sigma$ away from its SM prediction, depending on the value of $V_{cb}$ used to calculate its associated theoretical prediction. 

Finally, concerning LFU observables, the LHCb collaboration recently released new results for the ratios $R_{K^{(*)}}$ in four bins of the dilepton invariant mass, utilising 9fb$^{-1}$ of data~\cite{LHCb:2022zom,LHCb:2022qnv},
\begin{align}
\begin{aligned}
{R_K}^{[0.1,1.1]}_\text{LHCb} &= 0.994^{+0.094}_{-0.087} \;\; (-0.0\sigma),\\
{R_K}^{[1.1,6]}_\text{LHCb} &= 0.949^{+0.048}_{-0.047} \;\; (+1.1\sigma),\\
{R_{K^*}}^{[0.1,1.1]}_\text{LHCb} &= 0.927^{+0.099}_{-0.093} \;\; (+0.5\sigma),\\
{R_{K^*}}^{[1.1,6]}_\text{LHCb} &= 1.027^{+0.077}_{-0.073} \;\; (-0.4\sigma).
\end{aligned}
\end{align}
All four measurements show, at present, no significant deviation from the SM expectations, contrary to earlier LHCb measurements, which had indicated the violation of LFU. Therefore, the NP contribution should be dominantly LFU.

\subsection{Global Model-Independent Fits}\label{globalfits}

The effective Hamiltonian, valid below the EW scale, relevant for $b\to s\ell^+\ell^-$, can be written as~\cite{Grinstein:1987vj,Buchalla:1995vs}
\begin{equation}\label{eq:WETHamiltonian}
\mathcal{H}_\text{eff} = -\frac{4G_F}{\sqrt{2}} V_{tb}V_{ts}^* \sum_{i} C_i(\mu) \mathcal{O}_i(\mu),
\end{equation}
where $G_F$ is the Fermi constant, $V_{tb}$ and $V_{ts}$ are CKM matrix elements, $\mathcal{C}_i(\mu)$ are the Wilson coefficients that encode the short-distance dynamics, and $\mathcal{O}_i(\mu)$ are the corresponding effective operators. The scale $\mu$ represents the renormalisation scale at which the Hamiltonian is evaluated. In addition to the SM operators, the effective Hamiltonian may also include operators that encapsulate structures not generated in the SM, i.e.~right-handed currents or scalar interactions, arising in various NP scenarios. The most relevant operators for the following discussion are
\begin{eqnarray}
\begin{aligned}
\mathcal{O}_7^{(\prime)} &=& (\bar{s}\sigma_{\mu\nu}P_{R(L)}b)F^{\mu\nu},\\
{\mathcal{O}}_{9\ell}^{(\prime)} &=& (\bar{s} \gamma_{\mu} P_{L(R)} b) (\bar{\ell} \gamma^\mu \ell),\\
{\mathcal{O}}_{10\ell}^{(\prime)} &=& (\bar{s} \gamma_{\mu} P_{L(R)} b) (\bar{\ell} \gamma^\mu \gamma_5 \ell),
\label{bslloperators}
\end{aligned}
\end{eqnarray}
where colour indices have been actively omitted, $P_{L,R} = (1 \mp \gamma_5)/2$ are the chirality projection operators, $F^{\mu\nu}\equiv \partial^\mu A^\nu-\partial^\nu A^\mu$ is the electromagnetic-field strength tensor ($A^\mu(x)$ being the photon field), $\sigma^{\mu\nu}=\frac{i}{2}[\gamma^\mu,\gamma^\nu]$ with $\gamma^\mu$ the gamma matrices in four dimensions, and $b,s$ denote the quark fields.

The Wilson coefficients corresponding to the most relevant operators of the effective Hamiltonian in Eq.~\ref{eq:WETHamiltonian} present the following values within the SM at the scale $\mu_b=4.8$ GeV~\cite{Gambino:2003zm,Bobeth:2003at,Misiak:2006ab,Huber:2005ig,Huber:2007vv}:
\begin{align}
\begin{aligned}
{\cal C}_7^\text{eff}(\mu_b)&=-0.2923\,,\\ {\cal C}_9(\mu_b)&=+4.0749\,,\\
{\cal C}_{10}(\mu_b)&=-4.3085\,.
\end{aligned}
\end{align}
Several groups have performed global fits to different combinations of Wilson Coefficients~\cite{Alguero:2023jeh,Greljo:2022jac,Ciuchini:2022wbq,Hurth:2021nsi}, including 1D, 2D, and multidimensional scenarios with up to 20 independent Wilson coefficients simultaneously. Notably, the NP scenario $\mathcal{C}_{9}^{\rm U}$, representing a lepton-flavour universal contribution~\cite{Alguero:2018nvb}, i.e.~$\mathcal{C}_{9\mu}=\mathcal{C}_{9e}\equiv\mathcal{C}_{9}^{\rm U}$, is particularly effective in explaining the observed deviations\footnote{Interestingly it was recently found in Ref.~\cite{Isidori:2023unk} that semi-inclusive $b \to s \ell^+\ell^-$ transitions at high-q$^2$ also points towards the same solution.}, especially in light of the new LHCb measurements of $R_{K(^*)}$ and the CMS measurement of $B_s\to\mu^+\mu^-$. In this section, we will primarily focus on the results presented in Ref.~\cite{Alguero:2023jeh}, and later do a comparison with the results of other groups.

\begin{table*}[!ht]
\centering
\begin{adjustbox}{width=0.8\textwidth, center=\textwidth}
\begin{tabular}{lc||c|c|c|c}
\multicolumn{2}{c||}{Scenario} & Best-fit point & 1$\sigma$ & Pull$_{\rm SM}$ & p-value \\
\hline\hline
$b\to s\ell^+\ell^-$ & $\mathcal{C}_{9}^{\rm U}$ & $-1.17$ & $[-1.33,-1.00]$ & 
5.8 & 39.9 \%\, \\
\hline
\multirow{ 2}{*}{$b\to s\ell^+\ell^-$} & $\mathcal{C}_{9}^{\rm U}$ & $-1.18$ & $[-1.35, -1.00]$ &
\multirow{ 2}{*}{5.5} & \multirow{2}{*}{39.1\,\%} \\
&$\mathcal{C}_{10}^{\rm U}$ & $+0.10$ & $[-0.04, +0.23]$ &\\
\hline
\multirow{ 2}{*}{$b\to s\ell^+\ell^-$} & $\mathcal{C}_{9\mu}^{\rm V}=-\mathcal{C}_{10\mu}^{\rm V}$ & $-0.08$ & $[-0.14, -0.02]$ &
\multirow{ 2}{*}{5.6} & \multirow{2}{*}{41.1\,\%} \\
&$\mathcal{C}_{9}^{\rm U}$ & $-1.10$ & $[-1.27, -0.91]$ &\\
\hline\hline
\multirow{2}{*}{$b\to s\ell^+\ell^-+R({D^{(*)}})$} & $\mathcal{C}_{9\mu}^{\rm V}=-\mathcal{C}_{10\mu}^{\rm V}$ & $-0.11$ & $[-0.17, -0.05]$ &
\multirow{ 2}{*}{6.3} & \multirow{2}{*}{35.4\,\%} \\
&$\mathcal{C}_{9}^{\rm U}$ & $-0.78$ & $[-0.90, -0.66]$ &\\
\hline
\end{tabular}
\end{adjustbox}
\caption{Most prominent scenarios that emerge from a global fit to $b\to s\ell^+\ell^-$ data (see also \cite{Alguero:2023jeh,Alguero:2019ptt}).
}\label{table:Fitsbsll} 
\end{table*}

\begin{figure*}[ht]
\centering
\includegraphics[width=7.0 cm]{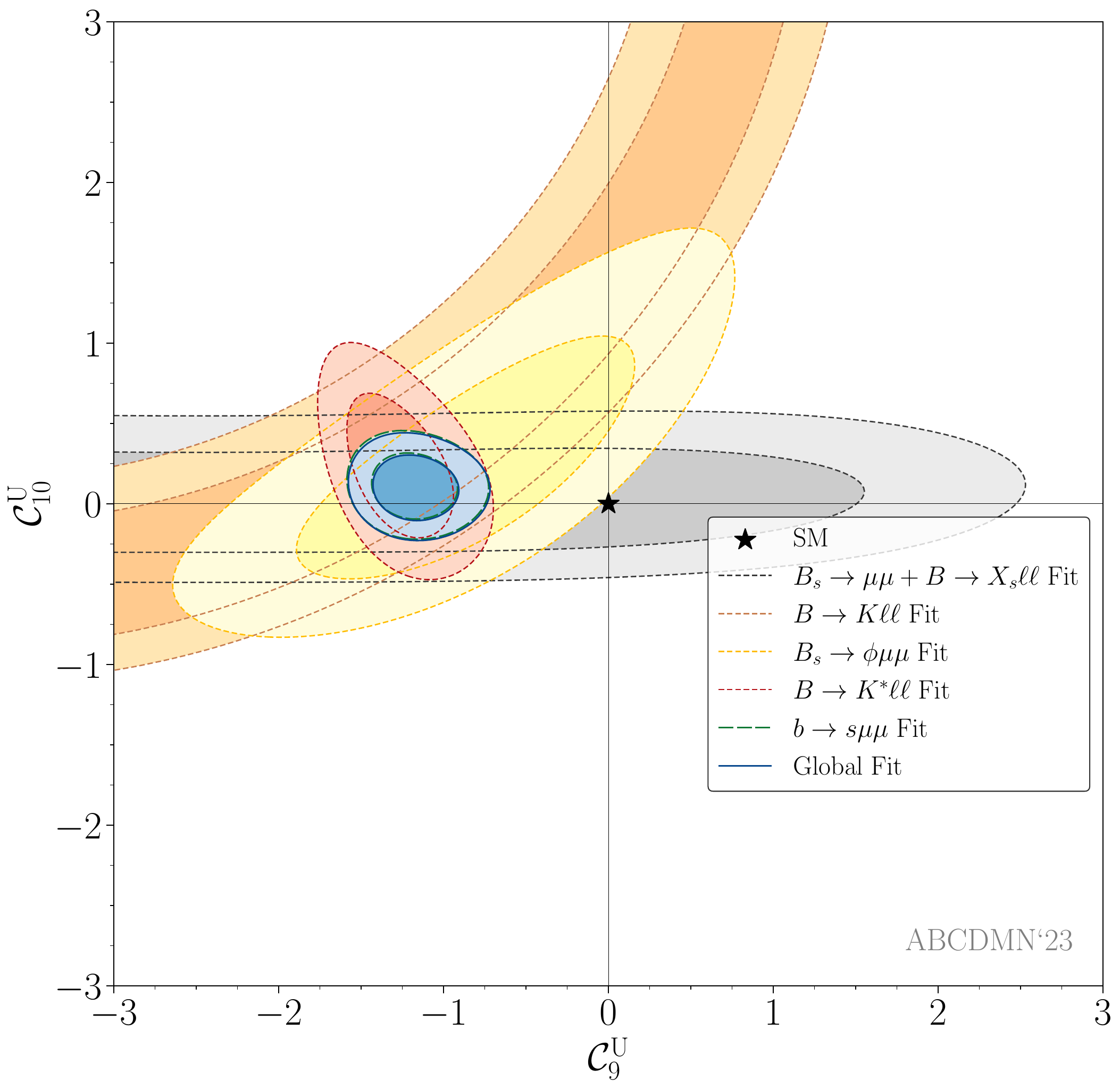} \quad
\includegraphics[width=7.0 cm]{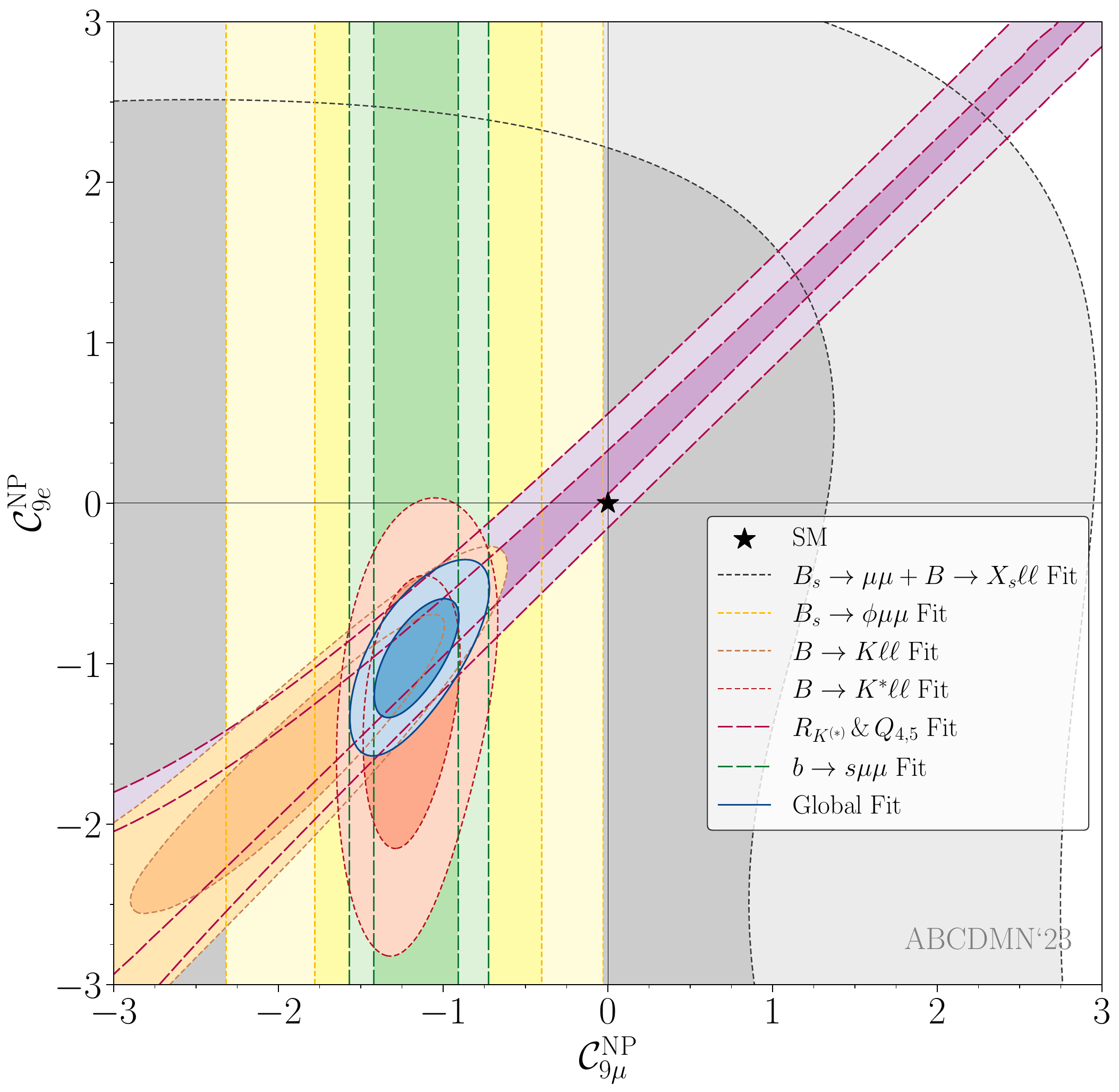}
\caption{1$\sigma$ (dark-shaded) and 2$\sigma$ (light-shaded) confidence regions for $(\mathcal{C}_{9}^{\rm U},\mathcal{C}_{10}^{\rm U})$ (left) and $(\mathcal{C}_{9\mu}^{\rm NP},\mathcal{C}_{9e}^{\rm NP})$ scenarios (right). Distinct fits are performed separating each of the $b\to s\ell^+\ell^-$ modes (short-dashed contours), the LFUV observables and the  combined $b\to s\mu^+\mu^-$ modes (long-dashed contours), and the global fit (solid contours). The colour code is provided in the individual captions.  Notice that some fits (for instance the $B\to K^{(*)}\ell^+\ell^-$ Fit(s) and the LFUV Fit) share a number of observables and thus are not completely uncorrelated.}\label{fig:C9U_C10U_C9mu_C9e}
\end{figure*}

A selection of the preferred NP scenarios can be found in Table~\ref{table:Fitsbsll}. The confidence region plots in Fig.~\ref{fig:C9U_C10U_C9mu_C9e} depict the 1$\sigma$ and 2$\sigma$ contours for the 2D scenarios $(\mathcal{C}_{9}^{\rm U},\mathcal{C}_{10}^{\rm U})$ and $(\mathcal{C}_{9\mu}^{\rm NP},\mathcal{C}_{9e}^{\rm NP})$, showing regions constrained by the several individual modes that constitute the global fits, LFUV observables, only $b\to s\mu^+\mu^-$ modes, and the global fit.

For the $(\mathcal{C}_{9}^{\rm U},\mathcal{C}_{10}^{\rm U})$ scenario, the grey contour, which mainly reflects the constraints posed by $\mathcal{B}_{B_s\to\mu^+\mu^-}$ and $\mathcal{B}_{B\to X_s\ell^+\ell^-}$, primarily indicates consistency with $\mathcal{C}_{10}^{\rm U} = 0$, although it could accommodate small NP contributions to $\mathcal{C}_{10}^{\rm U}$ with a positive sign. This is also the case for all the other modes included in the global fit. The main reason behind this behaviour can be found in the fact that the current global average of ${\cal B}_{B_s\to\mu^+\mu^-}$ aligns with the corresponding SM estimate at the level of $\sim 1\sigma$. On the other hand, looking at the horizontal axis, all the constraints are consistent within 1$\sigma$ with a value of $\mathcal{C}_{9}^{\rm U} \sim -1$.

In the $(\mathcal{C}_{9\mu}^{\rm NP},\mathcal{C}_{9e}^{\rm NP})$ scenario, the impact of the new LHCb measurements of $R_{K^{(*)}}$ is evident, leading to an agreement among all the different pieces that constitute the global fit for a NP contribution suggesting $\mathcal{C}_{9\mu}^{\rm NP}=\mathcal{C}_{9e}^{\rm NP}$, which indicates a strong signal of LFU NP associated to the semileptonic $O_{9\ell}$ operator. The combination of $b\to s\mu^+\mu^-$ modes cannot place any bound on $\mathcal{C}_{9e}^{\rm NP}$, explaining why this region is unconstrained w.r.t.~this axis in the plot. $B\to K\ell^+\ell^-$ favours negative values for both $\mathcal{C}_{9\mu}^{\rm NP}$ and $\mathcal{C}_{9e}^{\rm NP}$ and are consistent with the relation $\mathcal{C}_{9\mu}^{\rm NP}=\mathcal{C}_{9e}^{\rm NP}$ at 1$\sigma$, mainly due to $R_K$ being the only $B\to K\ell^+\ell^-$ observable contributing to $\mathcal{C}_{9e}^{\rm NP}$. The $B\to K^{*}\ell^+\ell^-$ observables also prefer negative values for both Wilson coefficients, with negligible correlation. The same also applies to the $B_s\to \phi\ell^+\ell^-$ mode, which is also compatible with all the other ones at the level of $1\sigma$, but with larger errors.

\begin{figure}[ht] 
\begin{center}
\includegraphics[width=7.5 cm]{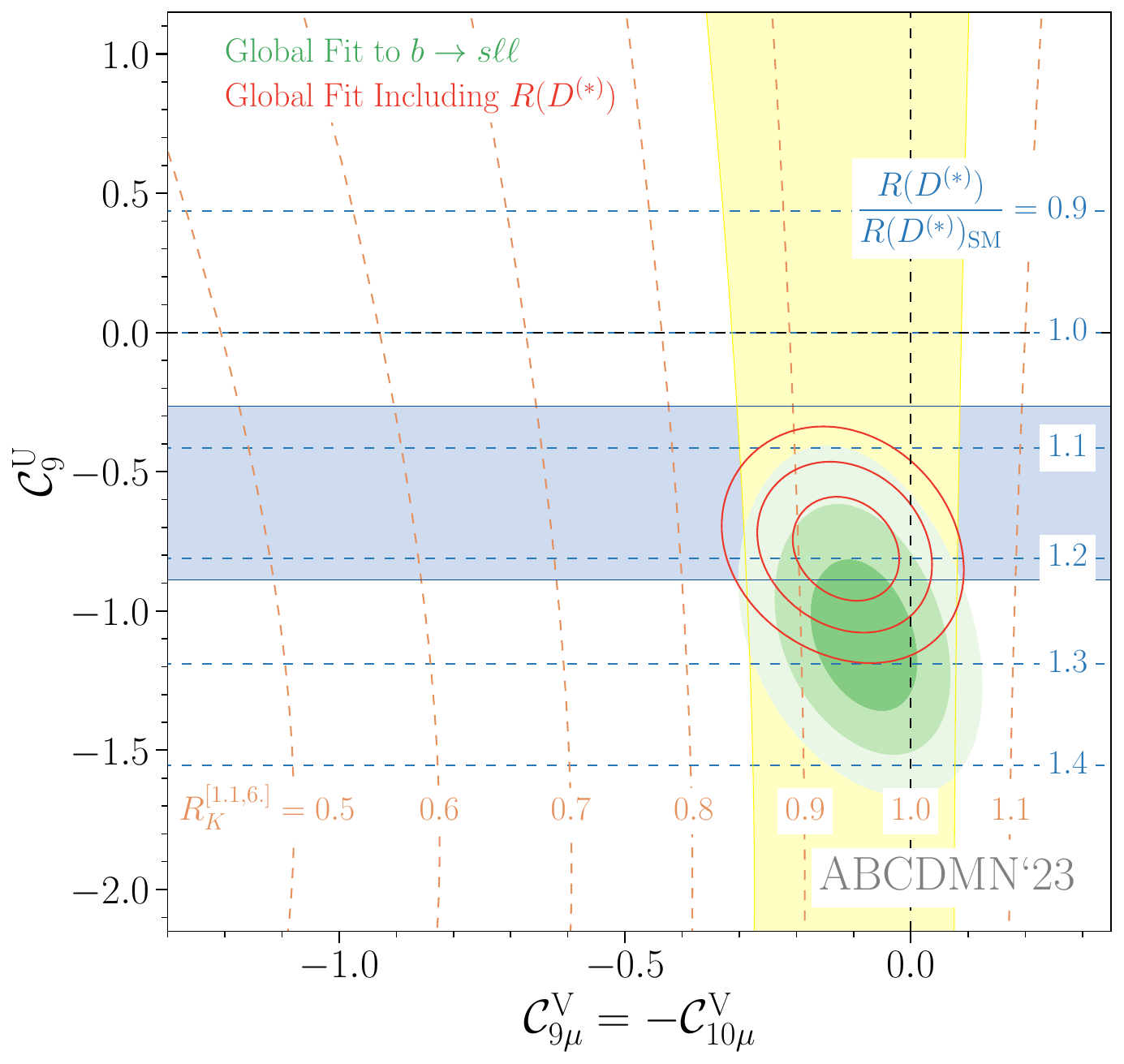}
\end{center}  
\caption{Preferred regions at the 1$\sigma$, 2$\sigma$ and 3$\sigma$ level (green) in the $(\mathcal{C}_{9\mu}^{\rm V}=-\mathcal{C}_{10\mu}^{\rm V},\mathcal{C}_{9}^{\rm U})$ plane from $b\to s\ell^+\ell^-$ data. The red contour lines show the corresponding regions once $R_{D^{(\ast)}}$ is included (for a NP scale $\Lambda=2$ TeV). The horizontal blue (vertical yellow) band is consistent with $R_{D^{(\ast)}}$ ($R_K$) at the 2$\sigma$ level and the contour lines show the predicted values for these ratios. }
\label{fig:Sc8RX}   
\end{figure}

The more complicated NP scenario $(\mathcal{C}_{9\mu}^{\rm V}=-\mathcal{C}_{10\mu}^{\rm V}, \mathcal{C}_{9}^{\rm U})$
~\cite{Alguero:2018nvb}
, apart from being one of the best from a quality-of-fit perspective, allows to establish a model-independent connection between charged and neutral anomalies as we will see in the next section. Figure~\ref{fig:Sc8RX} presents the corresponding preferred regions and the corresponding numerical values are given in Table.~\ref{table:Fitsbsll}.

Let us now compare the results of global $b\to s\ell^+\ell^-$ fits from various groups:
\begin{itemize}
\item ABCDMN: M.~Algueró, A.~Biswas, B.~Capdevila, S.~Descotes-Genon, J.~Matias, M.~Novoa-Brunet \cite{Alguero:2023jeh}.
\item AS/GSSS: W.~Altmannshofer, P.~Stangl / A.~Greljo, J.~Salko, A.~Smolkovic, P.~Stang \cite{Greljo:2022jac}.
\item CFFPSV: M.~Ciuchini, M.~Fedele, E.~Franco, A.~Paul, L.~Silvestrini, M.~Valli \cite{Ciuchini:2022wbq}.
\item HMMN: T.~Hurth, F.~Mahmoudi, D.~Martínez-Santos, S.~Neshatpour \cite{Hurth:2021nsi}.
\end{itemize}
These collaborations use different statistical methods, FF choices and assumptions about the non-perturbative effects. For a detailed review of the different approaches, the readers are referred to Refs.~\cite{London:2021lfn,Albrecht:2021tul}. CFFPSV uses two different methods, the so-called Phenomenological Model Driven (PMD) and Phenomenological Data Driven (PDD) approaches. The PMD approach leverages existing LCSR estimates for the non-local FF to constrain their proposed polynomial parametrisation in $q^2$; they then adjust the parameters of this parametrisation to the $B\to K^*\mu^+\mu^-$ angular distributions while adhering to these constraints. Conversely, the PDD approach allows all parameters of their polynomial parametrisation to float freely without constraint, and fit them to the available data.

\begin{figure*}[ht]
\centering
\includegraphics[width=7.0 cm]{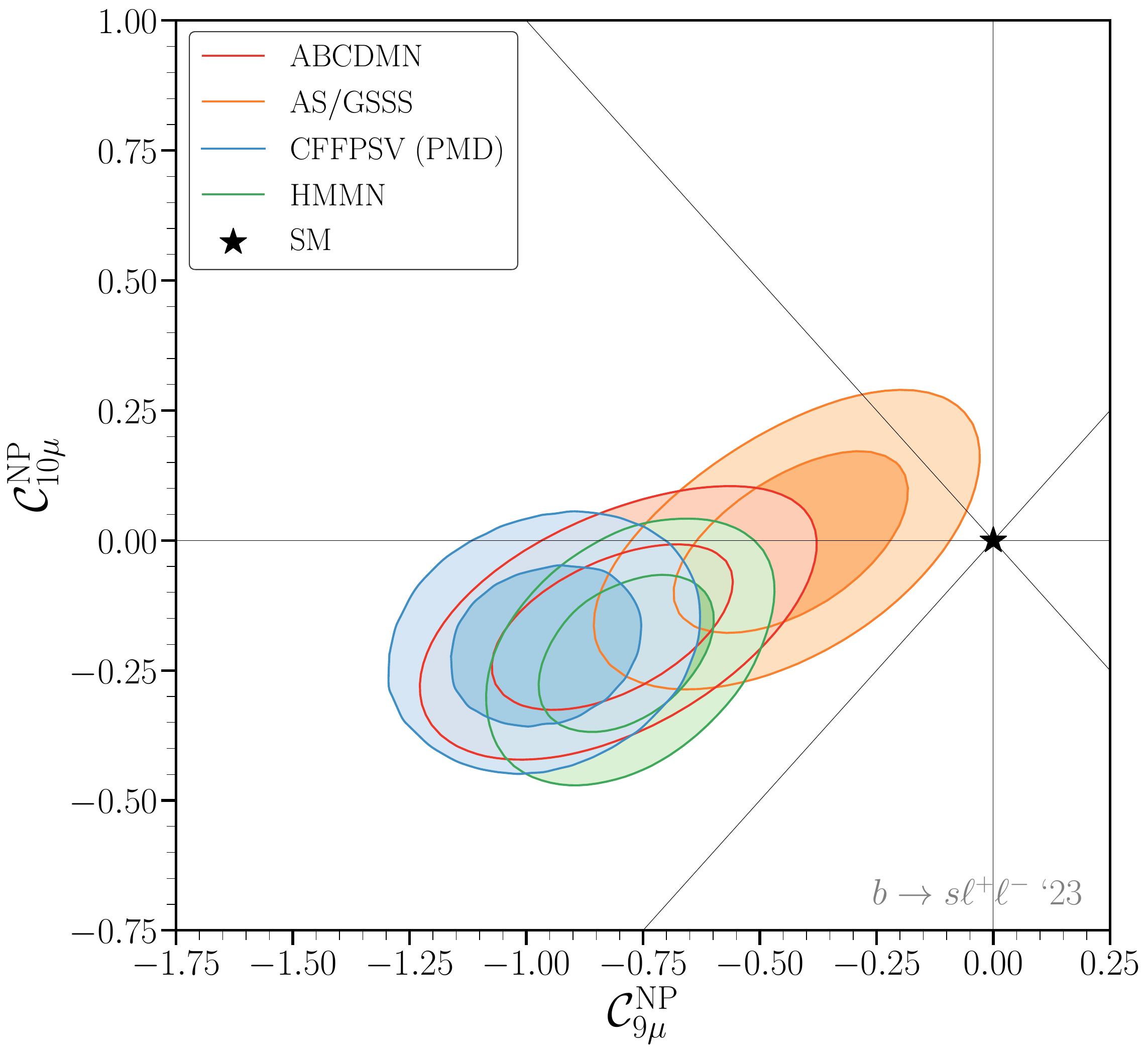}
\quad
\includegraphics[width=6.8 cm]{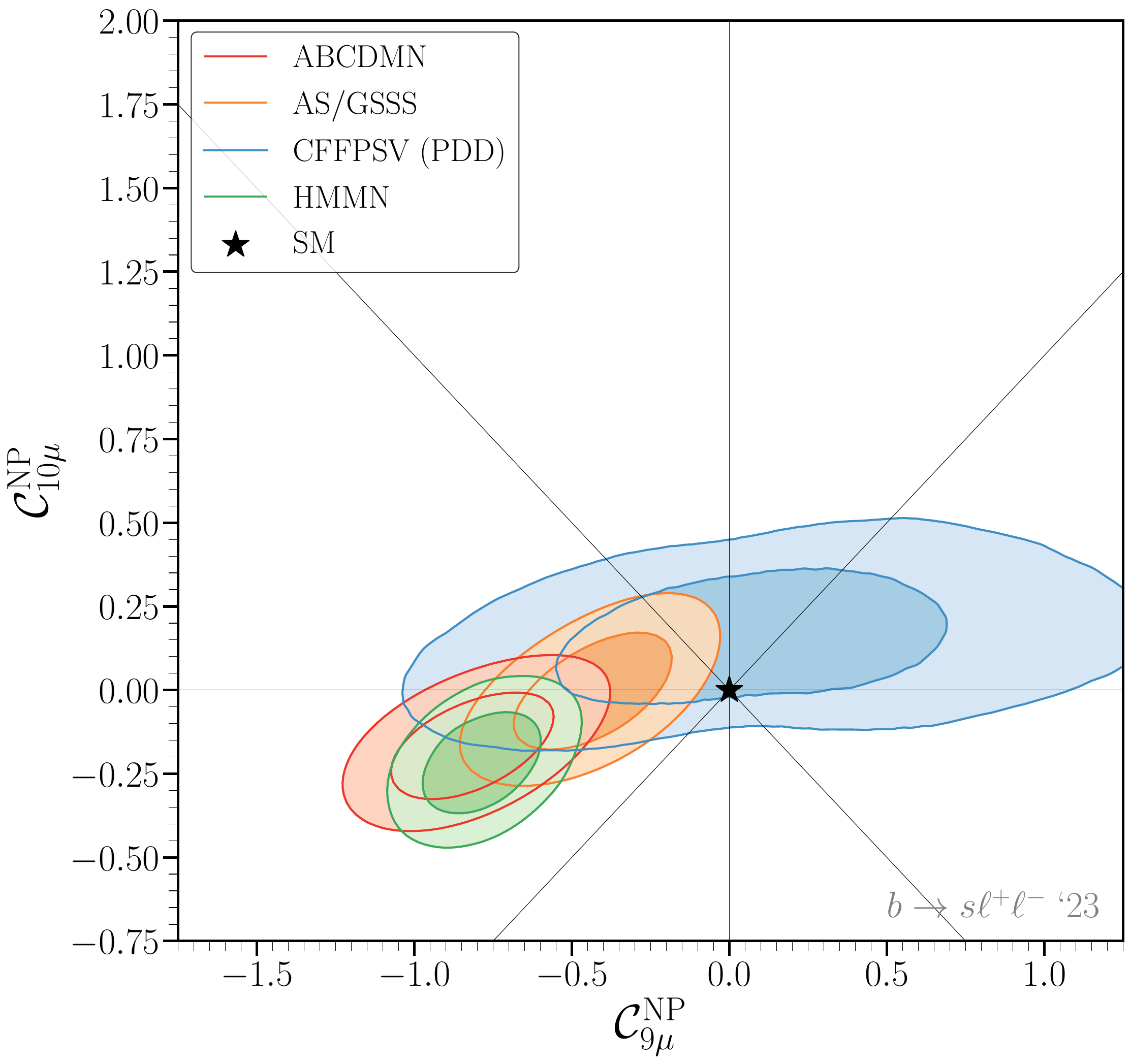}
\caption{Comparison of the results of the global fits of the different collaborations in the $C_9^\mu$-$C_{10}^\mu$ plane. Note that assuming LFU, the differences would be more pronounced and the significance for NP would be higher. We show the LFV case anyway because the data for comparison is available.}\label{fig:C9muNP_C10muNP_collabs}
\end{figure*}

Notably, as can be seen in Fig.~\ref{fig:C9muNP_C10muNP_collabs}, despite the diverse methodologies pursued, a good level of agreement is observed comparing the results in the $(\mathcal{C}_{9\mu}^\mathrm{NP},\mathcal{C}_{10\mu}^\mathrm{NP})$ plane. This convergence underscores the robustness and maturity attained in analysing $b\to s\ell^+\ell^-$ data, a pivotal conclusion of this review. However, there is one approach whose results disagree significantly with the others, namely the PDD approach from the CFFPSV group. This arises due to the very large number of free parameters that allow to fully absorb any potential effect. While the strategies adopted by the ABCDMN and HMMN collaborations share substantial similarities in terms of including all data, AS/GSSS does not incorporate measurements within the kinematic regime where $q^2>6$ $\mathrm{GeV}^2$. The inclusion of such data would harmonise their results with those of both the ABCDMN and HMMN collaborations. Concerning the latter two, the level of agreement and consistency is high, attributed to their similar data selection and treatment of non-perturbative effects, with the distinction that the ABCDMN collaboration solely focuses on meson decays while HMMN also include the baryonic decay $\Lambda_b\to\Lambda\mu^+\mu^-$.


While not included in Fig.~\ref{fig:C9muNP_C10muNP_collabs}, we should also briefly mention the results of the GRvDV (N.~Gubernari, M.~Reboud, D.~van Dyk, J.~Virto) group~\cite{Gubernari:2022hxn}. Their framework is based on a simultaneous Bayesian fit of all non-perturbative parameters, including the coefficients of the parameterisations for both local and non-local FFs, together with the NP contributions to the relevant Wilson Coefficients. In order to control the uncertainty on the parametrisation of the non-local FFs, GRvDV derived unitarity bounds for the corresponding parameters from the analytic structure of the EW current-$\mathcal{O}_{1,2}$ correlator underlying the non-local FFs. Due to the technical complexity of their framework, having to fit a large number of free parameters, only dedicated fits to $B\to K^*\mu^+\mu^-$, $B\to K\mu^+\mu^-$ and $B_s\to \phi\mu^+\mu^-$ are presented in Ref.~\cite{Gubernari:2022hxn}. 

\section{$b\to c\tau\nu$}\label{bctaunu}

\begin{figure*}[t]
\centering
 \includegraphics[width=0.7\linewidth]{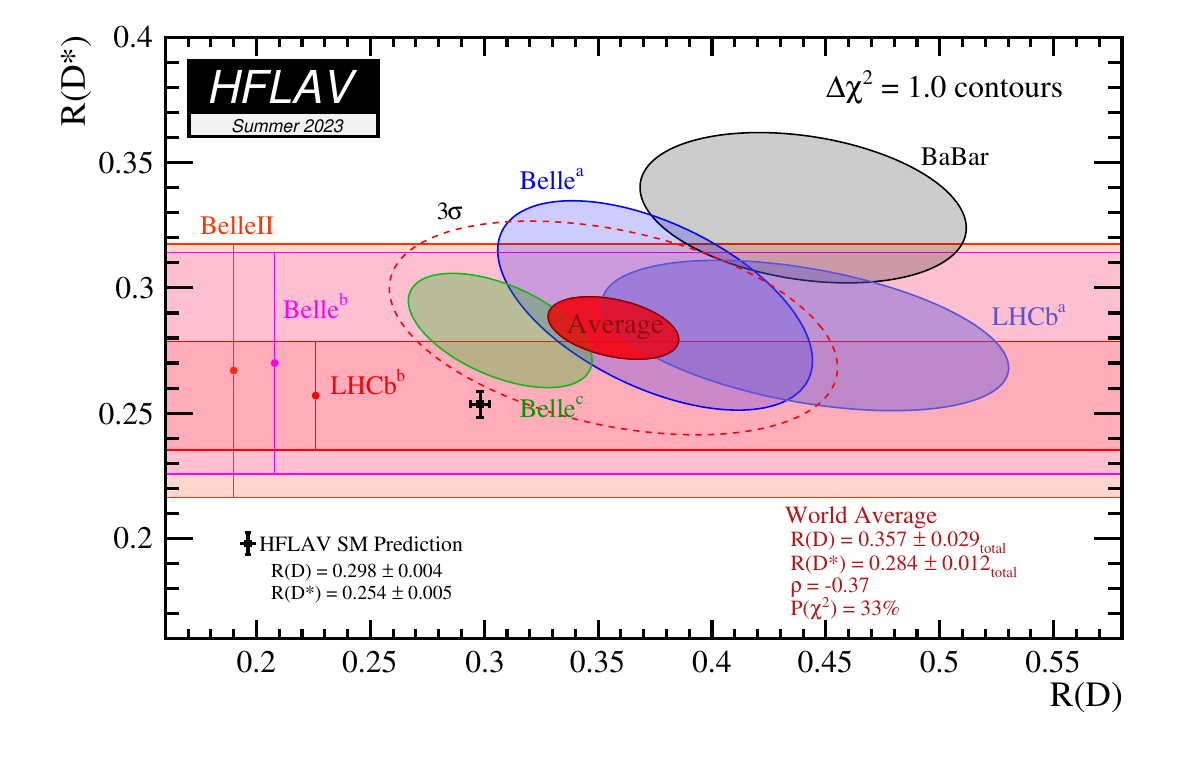}
 \caption{Summary of the measurements of $R(D)$ and $R(D^*)$ (including correlations) together with the SM predictions and according to the online update of HFLAV~\cite{HFLAV:2022pwe}.}
 \label{fig:RD}
\end{figure*}

The charged current $b\to c\tau\nu$ is already mediated at tree-level in the SM. Therefore, the corresponding decays have significant branching ratios, up to ${\mathcal O}(10^{-2})$. For light leptons (muons vs electrons) LFU is satisfied at the level of a few percent, i.e.~${R}_{e\mu}^{D^*} = {\cal B}_{B\to D^{*} e \bar\nu}/{\cal B}_{B\to D^{*} \mu \bar\nu} = 0.990\pm0.031$~\cite{Belle:2023bwv} and ${R}_{e\mu}^{D^*}=1.001\pm0.023$~\cite{BelleIISem}. However, the ratios
\begin{equation}
{R}({D^{\left(*\right)}}) = {{\cal B}_{B \to {D^{\left(*\right)}}\tau\nu} / {\cal B}_{{B\to{D^{\left(*\right)}}\ell\nu }}}\,,
\end{equation}
with $\ell=e,\mu$ are the main drivers of the anomalies. The tension between the average of the measurements of BaBar~\cite{BaBar:2012obs,BaBar:2013mob}, Belle~\cite{Belle:2015qfa,Belle:2016ure,Hirose:2016wfn,Hirose:2017dxl, Belle:2019rba} and LHCb~\cite{Aaij:2015yra,Aaij:2017uff,Aaij:2017deq,LHCb:2023uiv}
\begin{eqnarray}
\begin{aligned}
{R}(D)\,=\,0.356\pm0.029\,,\\
{R}(D^*)\,=\,0.284\pm0.013\,,
\label{eq:HFLAV}
\end{aligned}
\end{eqnarray}
and the corresponding SM predictions
 \begin{eqnarray}
\begin{aligned}
{R}_{\rm SM}(D)\,=\,0.298\pm0.004 \,, \\
{R}_{\rm SM}(D^*) \,=\,0.254\pm0.005 \,,
\label{eq:HFLAVSM}
\end{aligned} 
\end{eqnarray}
amounts to $3.2\,\sigma$~\cite{HFLAV:2022pwe}. 
This situation is illustrated in Fig.~\ref{fig:RD}.

An analogous behavior, i.e.~and enhancement w.r.t.~the SM prediction, has been observed for ${\cal R}(J/\psi)={\cal B}_{B_c\to J/\psi\tau\nu}/{\cal B}_{B_c\to J/\psi\mu\nu}$ with the measurement~\cite{LHCb:2017vlu}
\begin{equation}
    {R}(J/\psi)=0.71 \pm 0.17 \pm 0.18\,,
\end{equation}
and the SM prediction~\cite{Cohen:2018dgz,Leljak:2019eyw,Harrison:2020nrv,Harrison:2020gvo}
\begin{equation}
 {R}_{\rm SM}(J/\psi) = 0.258 \pm 0.004\,.   
\end{equation}
However, LHCb~\cite{LHCb:2022piu} finds
\begin{equation}
{R}(\Lambda_c) = 0.242 \pm 0.026 \pm 0.040 \pm 0.059\,,
\end{equation}
where the first uncertainty is statistical, the second is systematic and the third is due to external branching fraction measurements. Concerning the SM prediction, a recent reanalysis~\cite{Bernlochner:2022hyz} yields\footnote{The SM prediction, where the absence of a subleading Isgur-Wise function at $\mathcal{O}(\bar \Lambda/m_{c,b})$ in the $\Lambda_b \to \Lambda_c$ transition suppresses the theoretical uncertainty~\cite{Neubert:1993mb}, is equal to~\cite{Gutsche:2015mxa,Shivashankara:2015cta,Detmold:2015aaa,Li:2016pdv,Datta:2017aue,Bernlochner:2018kxh,Bernlochner:2018bfn} ${\cal R}_{\rm SM}(\Lambda_c) = 0.324 \pm 0.004$}
\begin{equation}
R_{\rm SM}(\Lambda_c) = |0.04/V_{cb}|^2 (0.285 \pm 0.073)\,.
\end{equation}
This means that in a global fit, the effects of including ${R}(\Lambda_c)$ and ${R}(\Jpsi)$ tend to cancel and, in fact, they are often disregarded in global fits. Furthermore, it can be shown even if generic NP is included, ${R}(\Lambda_c)$ is in tension with ${R}(D^{(*)})$ due to a general sum-rule~\cite{Blanke:2018yud,Blanke:2019qrx,Fedele:2022iib}.\footnote{A PhD thesis on $R(D^{(*)})$ analyzing BaBar data exists which finds significantly lower values~\cite{Li:2022amw}. However, these results are neither published nor approved by the BaBar collaboration. }

Here, a comment on the SM prediction is in order. Unlike $R_{K^{(*)}}$, ${R}(D^{(*)})$ possess imperfect cancellations of the FF dependence since the tau mass is sizeable compared to the $B$ mass. This means that while the overall normalization drops out, one is sensitive to the shape of the FFs. While the HFLAV SM prediction is based on FFs including data from $B\to D^{(*)}\ell\nu$ with light leptons in order to reduce the uncertainties~\cite{Bigi:2016mdz,Bernlochner:2017jka,Jaiswal:2017rve,Gambino:2019sif,Bordone:2019vic,Martinelli:2021onb}, there are determinations that rely on lattice QCD but do not include experimental $B\to D^{(*)}\ell\nu$ input~\cite{FermilabLattice:2021cdg,Harrison:2023dzh}. Here, the Dispersive Matrix approach~\cite{DiCarlo:2021dzg,Martinelli:2021frl} stands out, since it reduces the tension in ${R}_{\rm SM}(D^{*})$ to $1.3\sigma$~\cite{Martinelli:2021myh}. However, this comes at the cost of significant tensions between the SM predictions and the measurements of polarization observables, in particular, the $D^*$ longitudinal polarization fraction $F_L^{\ell}$~\cite{Belle:2023bwv,BelleIISem}, at the $\sim\!4\sigma$ level~\cite{Fedele:2023ewe}. While this of course assumes the absence of NP in light leptons, which is not guaranteed~\cite{Carvunis:2021dss} despite the agreement in total muon vs electron rates, it can be shown that even completely generic NP cannot significantly reduce the tension in $F_L^{\ell}$~\cite{Fedele:2023ewe}.

\subsection{$b\to c\tau\nu$ Model Independent Results}

The effective Hamiltonian governing $b\to c\tau\nu$ transitions is usually defined as
\begin{equation}
\renewcommand{\arraystretch}{1.8}
\begin{array}{r}\label{eq:Heff}
 {\cal H}_{\rm eff}=  2\sqrt{2} G_{F} V^{}_{cb} \big[(1+g_{V_L}^{\ell}) O_{V_L}^{\ell} + g_{V_R}^{\ell} O_{V_R}^{\ell}
 \\   + g_{S_L}^{\ell} O_{S_L}^{\ell} + g_{S_R}^{\ell} O_{S_R}^{\ell}+g_{T}^{\ell} O_{T}^{\ell}\big] + \text{h.c.}\,,
\end{array}
\end{equation}
with the dimension-six operators\footnote{This does not include the case of light right-handed neutrinos.}
\begin{equation}
\renewcommand{\arraystretch}{1.8}
\begin{array}{l}
   O_{V_L}^{\ell}  = \left(\bar c\gamma ^{\mu } P_L b\right)  \left(\bar {\ell} \gamma_{\mu } P_L \nu_{{\ell}}\right)\,, \\ 
   O_{V_R}^{\ell}  = \left(\bar c\gamma ^{\mu } P_R b\right)  \left(\bar {\ell} \gamma_{\mu } P_L \nu_{{\ell}}\right)\,, \\ 
   O_{S_L}^{\ell}  = \left( \bar c P_L b \right) \left( \bar {\ell} P_L \nu_{{\ell}}\right)\,,   \\
   O_{S_R}^{\ell}  = \left( \bar c P_R b \right) \left( \bar {\ell} P_L \nu_{\ell}\right)\,, \\
   O_{T}^{\ell}  = \left( \bar c \sigma^{\mu\nu}P_L  b \right) \left( \bar {\ell} \sigma_{\mu\nu} P_L \nu_{{\ell}}\right)\,.
\end{array}
\label{eq:Oeff}
\end{equation}
%
There are several scenarios that can accommodate data nicely. In particular, the quite recent analysis Ref.~\cite{Iguro:2022yzr} finds for the one-dimensional case
\begin{align}
&\begin{array}{ccc}
 &{\rm Pull} & {\rm Best}\;{\rm Fit}\;{\rm Point}\\
\hline C_{V_L} & 4.4\sigma & +0.08(2) \\
C_{V_R} & 1.9\sigma& -0.05(3) \\
C_{S_L} & 3.0\sigma & +0.17(5) \\
C_{S_R} & 3.8\sigma& +0.20(5) \\
C_T & 3.4\sigma & -0.03(1)
\end{array}
\end{align}
assuming real Wilson coefficients. Note that here the theoretical uncertainties stemming from FFs were neglected in the NP contribution such that including them would slightly reduce the significance. Also note that at the dimension-six level in the SMEFT, $g_{V_R}$ is lepton flavour-universal, implying $g_{V_R}^e=g_{V_R}^\mu=g_{V_R}^\tau$, i.e.~no effect in $R(D^{(*)})$. The fit can be further improved by allowing for complex Wilson coefficients. Note that in general the best-fit points are still allowed by the $B_c$ lifetime as LHC constraints. Finally, there are the following 2-dimensional LQ-motivated scenarios~\cite{Iguro:2022yzr} 
\begin{equation}
\begin{array}{*{20}{l}}
{{\rm{LQ}}}& {{\rm Best}\;{\rm Fit}\;{\rm Point}}\\
\hline
{{U_1}}& {{C_{{V_L}}} = 0.07,{C_{{S_R}}} = 0.06}\\
{{{\rm{S}}_1}}& {{C_{{V_L}}} = 0.06,{C_{{S_L}}} =  - 8.9{C_T} = 0.06}\\
{{{\rm{R}}_2}}& {{C_{{V_R}}} =  \pm i0.68,{C_{{S_L}}} = 8.4{C_T} = 0.04 \mp i0.65}\,,
\end{array}
\end{equation}
all with pulls of $3.8\sigma$. 

\section{New Physics Explanations}\label{npexplanations}

\subsection{$b\to s\ell^+\ell^-$}
 
As these processes are suppressed in the SM, the required $O(20\%)$ NP effect (w.r.t.~the SM) is small and we have mainly two different classes of solutions which can provide naturally a dominantly LFU effect in $\mathcal{C}_9$. $Z^\prime$ boson can already contribute at tree-level~\cite{Buras:2013qja,Gauld:2013qba,Altmannshofer:2014cfa,Crivellin:2015mga}, however, $B_s-\bar B_s$ mixing constraints~\cite{DiLuzio:2017fdq}, LHC limits~\cite{Allanach:2015gkd} and bounds on 4-lepton contact interactions from LEP~\cite{ALEPH:2006bhb} must be respected. These constraints do not allow to account for the best fit value of $\mathcal{C}_9^{\rm U}$ in a simplified $Z^\prime$ model with only left-handed $\bar s b$ couplings~\cite{Greljo:2022jac}. However, these bounds can be avoided, or at least weakened, if one allows for a small right-handed $\bar s b$ couplings~\cite{Buras:2012jb,Crivellin:2015era}, or for a cancellation between the $Z^\prime$ and Higgs contribution in $B_s-\bar B_s$ mixing which arises from the flavour symmetry breaking~\cite{Crivellin:2015lwa}. Furthermore, $K^0-\bar K^0$ and $D^0-\bar D^0$ enforce an approximate global $U(2)$ flavour symmetry on the $Z^\prime$ couplings to quarks~\cite{Calibbi:2019lvs}.

Also operators with charm quarks~\cite{Jager:2017gal} or tau leptons can generate an effect in $\mathcal{C}_9^{\rm U}$ via an off-shell photon penguin~\cite{Bobeth:2014rda}. Possible UV completions for this setup are the $U_1$ LQ~\cite{Crivellin:2018yvo}, the combination of the $S_1$ and $S_3$ leptoquarks (LQs)~\cite{Crivellin:2019dwb}, the $S_2$ LQ~\cite{Crivellin:2022mff}, a 2HDM with enhanced charm couplings~\cite{Iguro:2017ysu,Iguro:2018qzf,Crivellin:2019dun} and di-quarks~\cite{Crivellin:2023saq}.

On the other hand, direct (tree-level) LQ contributions to $\bar s b\bar\ell\ell$ operators are not favoured (anymore): One would both need a tuning among the contributions of two different representations to cancel the effect in $\mathcal{C}_{10}$ while obtaining a dominant $\mathcal{C}_9$. Furthermore, avoiding LFU processes like $\mu\to e\gamma$ is difficult~\cite{Crivellin:2017dsk} unless multiple LQ generations are present~\cite{Crivellin:2022mff}. Similar problems occur in models with loop-effects involving box contributions with new heavy scalars and fermions~\cite{Gripaios:2015gra,Arnan:2016cpy,Grinstein:2018fgb,Arnan:2019uhr}. 
 
\subsection{$b\to c\tau\nu$}
 
As this transition is tree-level mediated in the SM, also a tree-level NP contribution is necessary to obtain the desired effect of 10\% w.r.t.~the SM (for heavy NP with perturbative couplings). As it is a charged current process, the only possibilities are charged Higgses~\cite{Crivellin:2012ye,Fajfer:2012jt,Celis:2012dk}, $W^\prime$~bosons~\cite{Bhattacharya:2014wla,Greljo:2015mma,Boucenna:2016qad,Carena:2018cow} (with or without right-handed neutrinos) or LQs~\cite{Sakaki:2013bfa,Bauer:2015knc,Freytsis:2015qca,Fajfer:2015ycq}. While $W^\prime$ bosons have in general problems with LHC searches~\cite{Bhattacharya:2014wla,Greljo:2015mma}, despite bounds from the $B_c$ lifetime~\cite{Celis:2016azn,Alonso:2016oyd,Blanke:2019qrx,Aebischer:2021ilm} and LHC bounds~\cite{Blanke:2022pjy}, charged Higgses generating $\mathcal{C}_{S_R}$ can give a reasonably good fit to data. However, as LQs can generate $\mathcal{C}_{V_L}$ they are the best option for a full explanation, despite non-trivial constraints from $B_s-\bar B_s$ mixing, $B\to K^*\nu\nu$ and LHC searches arise. In particular $B\to K^*\nu\nu$ is very constraining such that the $SU(2)_L$ singlet vector LQ $U_1$~\cite{Calibbi:2015kma,Barbieri:2016las,DiLuzio:2017vat,Calibbi:2017qbu,Bordone:2017bld,Blanke:2018sro,Crivellin:2018yvo} or the singlet-triplet model ($S_1+S_3$)~\cite{Crivellin:2017zlb,Crivellin:2019dwb,Gherardi:2020qhc}, which can avoid these bounds, are particularly promising\footnote{Recently, BELLE II presented results with an excess in $B\to K^*\nu\nu$~\cite{BptoKpnunuEPS}, which could be related to the $B$ anomalies discussed here. However, the differential distribution seems to prefer light NP.}. In fact, $\mathcal{C}_9^{\rm U}$ is given by
\begin{align}
\mathcal{C}_9^{\rm U}\simeq7.5 \left(1-\sqrt{\frac{R({D^{(*)}})}{R(D^{(*)})_{\rm SM}}}\right)\left(1+\dfrac{{\rm log}\frac{\Lambda^2}{1 {\rm TeV}^2}}{10.5}\right)\,,
\end{align}
assuming large flavour-violating (i.e.~non-aligned) couplings. Using the combination of $D$ and $D^*$ this leads to  $R(D^{(*)})_{\rm exp}/R (D^{(*)})_{{\rm SM}}=1.142 \pm 0.039$, that implies $\mathcal{C}_{9}^{\rm U} \simeq -0.58$, assuming a NP scale $\Lambda$ of 2 TeV~\cite{Capdevila:2017iqn,Alguero:2022wkd}.
	
\subsection{Combined Explanations}

While charged Higgses can explain both $b\to s\ell^+\ell^-$ and $R(D^{(*)})$ in a small region in parameter space without violating LHC bounds~\cite{Iguro:2023jju}, the connection is indirect in the sense that different couplings are involved. For the $U_1$ LQ~\cite{Fuentes-Martin:2019ign,Fuentes-Martin:2020hvc,FernandezNavarro:2022gst} or the $S_1+S_3$ model~\cite{Crivellin:2019dwb,Gherardi:2020det}, the connection is more direct. Here the same couplings generating the contribution to $R(D^{(*)})$ also give rise to $C_9^{\rm U}$ via the off-shell photon penguin (modulus CKM rotations). Furthermore, a direct LFU violating effect is possible such that in $b\to s\ell^+\ell^-$ the $(\mathcal{C}_9^{\rm U},\mathcal{C}_{9\mu}^{\rm V}=-\mathcal{C}_{10\mu}^{\rm V})$ scenario arises. The combined fit to $b\to s\ell^+\ell^-$ and $R(D^{(*)})$ is shown in Fig.~\ref{fig:Sc8RX} in the limit of large flavour violating couplings.

\section{Correlations and prospects}\label{correlations}

In this section, we will discuss possible correlations with other observables and anomalies.

\subsection{$b \to s \tau^+\tau^-$}

\begin{figure}[ht] 
\begin{center}
\includegraphics[width=6 cm]{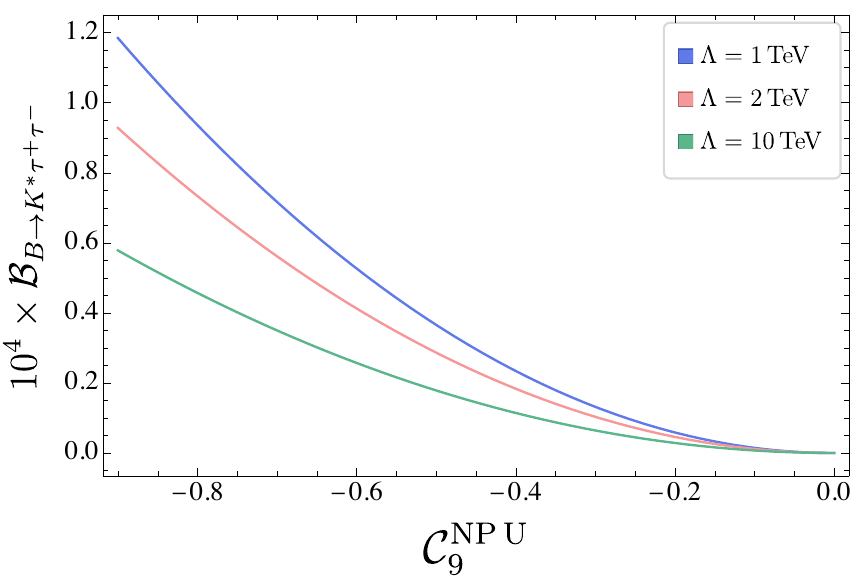}
\includegraphics[width=6.2 cm]{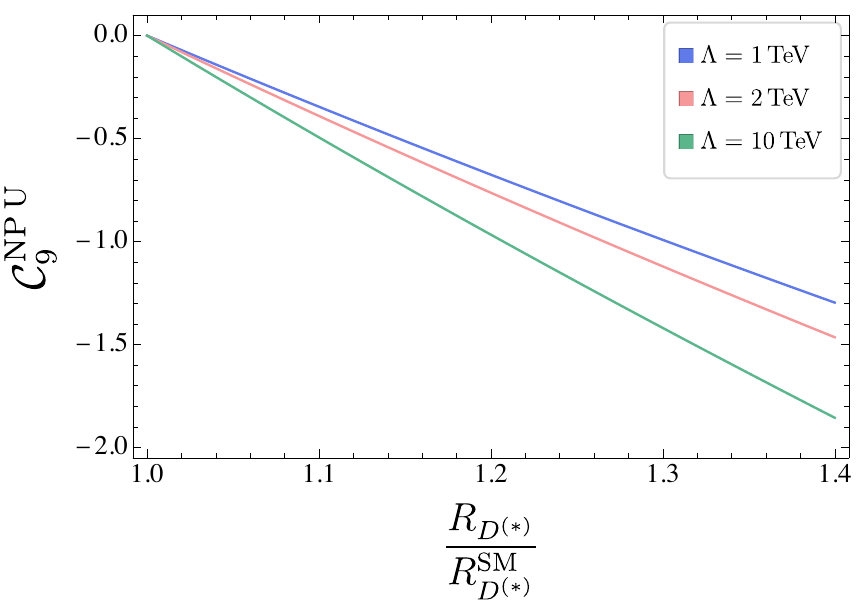}
\end{center}  
\caption{Predictions for Br$(B\to K^*\tau^+\tau^-)$ as a function of $\mathcal{C}_9^{\rm U}$ and $R(D^{(*)})/R(D^{(*)})_{\rm SM}$.}
\label{fig:C9URD}   
\end{figure}

If $R(D^{(*)})$ is explained via a left-handed vector current, not only a tau-loop contribution to $b\to s\ell^+\ell^-$ is generated via an off-shell photon penguin, but also $b\to s\tau^+\tau^-$ processes are significantly enhanced compared to their SM predictions. The reason for this is that due to $SU(2)_L$ invariance, $b\to c\tau \nu$ is related either to $b\to s\tau^+ \tau^-$ and/or $b\to s\nu_\tau \bar{\nu}_\tau$. However, the constraints on the latter from $B\to K^{(*)}\nu\nu$ are so stringent that the NP model must be chosen such that the effect is shifted to $b\to s\tau^+ \tau^-$, as is the case for the $U_1$ leptoquark and the $S_1+S_3$ combination. In fact, in this setup, one can predict the different $b\to s\tau^+ \tau^-$ as a function of $\mathcal{C}_9^{\rm U}$ or $R(D^{(*)})$ as shown in Fig.~\ref{fig:C9URD}, taken from Ref.~\cite{Alguero:2022wkd}. Note that if $\mathcal{C}_9^{\rm U}$ is tau-loop induced, this would also lead to a $q^2$ dependence measurable in the muon spectrum of $B\to K^{(*)}\mu^+\mu^-$\cite{Cornella:2020aoq}.

\subsection{Non-leptonic anomalies and connection to $R_{D^{(*)}}$}

One can construct (a different type of) optimised observable also in non-leptonic decays~\cite{Alguero:2020xca,Biswas:2023pyw} which benefits from cancellations on the theoretical and experimental side. In particular
\begin{eqnarray}
L_{K^*\bar{K}^*}&=&\rho(m_{K^*},m_{K^*})\frac{{\cal B}^{\rm long}_{\bar{B}_s \to K^* \bar{K}^*}}{{\cal B}^{\rm long}_{\bar{B}_d \to K^* \bar{K}^*}}
\nonumber \\[2mm]
L_{K\bar{K}}&=&
\rho(m_{K},m_{K})\frac{{\cal B}_{\bar{B}_s \to K \bar{K}}}{{\cal B}_{\bar{B}_d \to K \bar{K}}}
\end{eqnarray}
where the function $\rho$ is the inverse ratio of phase space factors of the two decays involved and can be taken to be one to a very good approximation (see \cite{Biswas:2023pyw}). However, the ratios $L_{K^*\bar{K}^*}$ and  $L_{K\bar{K}}$ contrary to the LFUV ones are controlled by U-spin symmetry that is obviously broken in the SM, but have the experimental advantage of involving only quarks (as electrons are highly problematic at LHCb). The SM prediction for these observables is~\cite{Biswas:2023pyw}:
\begin{equation}
L_{K^*\bar{K}^*}^{\rm SM}=19.53^{+9.14}_{-6.64} \quad  L_{K\bar{K}}^{\rm SM}=26.00^{+3.88}_{-3.59}\,,
\end{equation}
and the measured experimental values~\cite{ParticleDataGroup:2022pth,Belle:2012dmz,BaBar:2006enb,LHCb:2020wrt, Belle:2015gho,LHCb:2019bnl, BaBar:2007wwj}:
\begin{equation}
L_{K^*\bar{K}^*}^{\exp}=4.43\pm 0.92 \quad L_{K\bar{K}}=14.58\pm 3.37
\end{equation}
corresponding to a tension\footnote{A precise computation of the tension requires the use of the SM distribution of these observables that can be found in Ref.~\cite{Biswas:2023pyw}.} 
 with respect to the SM prediction of  2.6$\sigma$ and 2.4$\sigma$, respectively. The experimental uncertainty includes an extra 7\% relative uncertainty due to $B_s-\bar B_s$-mixing. 
They were computed using improved QCD-Factorization at NLO in $\alpha_s$ and the modelling of the infrared divergences followed the prescription of Ref.~\cite{Beneke:2006hg}.

In the EFT, it was pointed out in Ref.~\cite{Alguero:2020xca,Biswas:2023pyw}
that the same NP contribution to the Wilson coefficients ${\cal C}_{4q}$ of the QCD-penguin operator ${\cal O}_{4q^\prime}=
(\bar{q}_i^\prime b_j)_{V-A} \sum_q (\bar{q}_j q_i)_{V-A}
$ or ${\cal C}_{8gq}$ of the chromomagnetic operator ${\cal O}_{8gq}=-\frac{g_s}{8\pi^2} m_b \bar{q} \sigma_{\mu\nu}(1+\gamma_5)G^{\mu\nu} b$ provide  an explanation of both non-leptonic anomalies simultaneously. In the case of $L_{K \bar{K}}$ alone also the QCD-penguin operator ${\cal O}_{6q}$ can provide a solution. 

It was found in Ref.~\cite{Biswas:2023pyw} that the two ratios of mixed decays  ${\bar B}_s \to K^{*0} \bar{K}^0$
versus ${\bar B}_d \to \bar{K}^{*0} {K}^0$,   called $\hat{L}_{K^*}$, or ${\bar B}_s \to K^0 \bar{K}^{*0} $ 
versus ${\bar B}_d \to \bar{K}^{0} {K}^{*0}$,  called $\hat{L}_{K}$,
exhibit opposite behaviour, enhancement or decrease respectively, with respect to their rather close SM predictions under a NP contribution coming from the QCD-penguin ${\cal O}_{4q}$ or chromomagnetic operator ${\cal O}_{8gq}$.

Interestingly, one can also find model-dependent links~\cite{Lizana:2023kei} between observed tensions in optimized non-leptonic observables and the charged anomalies of $R_{D^{(*)}}$: The scalar leptoquark $S_1$ and a TeV-scale right-handed neutrino~\cite{Lizana:2023kei} are able to generate the necessary contribution to explain both non-leptonic anomalies. The model has two important features, it achieves a quasi-perfect cancellation of the contribution entering the electromagnetic dipoles ${\cal C}_{7\gamma s(d)}$ due to the hypercharge of the $S_1$ that induces naturally the cancellation in the RGE mixing of both dipoles. The scalar leptoquark $S_1$ is promoted to a doublet of $U(2)_q$ to avoid light-family flavour constraints and, finally, and more interesting the $S_1$ is a well-known solution of the charged-anomalies $R_{D^{(*)}}$ establishing a natural link between both type of anomalies and predicting a value for ${\cal B}_{B \to K \nu \bar{\nu}}$ very close to the present bound.
	
\section{Conclusions}\label{conclusions}

Semi-leptonic $B$ decays are important probes of the SM. They have in general controllable theory uncertainties, quite clear experimental signatures and are sensitive to NP effect due to their suppressed rates. We reviewed here the status of the anomalies, i.e.~deviations from the SM predictions, in $b\to s\ell^+\ell^-$ and $b\to c\tau\nu$ transitions.

The $b\to s\ell^+\ell^-$ transition is a favour-changing neutral current that is only mediated at the loop-level in the SM and is thus sensitive to small NP effects. Here the main drivers of the anomalies are~\cite{Alguero:2023jeh}: 
\begin{itemize}
    \item[\textit{i})] $\mathcal{B}^{\rm exp}_{B^+\to K^+\mu^+\mu^-}$ which deviates in several bins by 4$\sigma$ as well as a systematic trend in all other branching ratios albeit much less significant.
    \item[\textit{ii})] $P_5^\prime$: the most persistent tension that deviates by $\sim 2\sigma$ (considering only the neutral channel) in the two anomalous bins, supported by measurements of the charged mode.
    \item[\textit{iii})] Total branching ratio and angular observables in $B_s\to\phi\mu^+\mu^-$ depending on the choice of FFs used.
\end{itemize}
Importantly, these deviations from the SM form a consistent picture in the sense that a simple NP scenario can explain all discrepancies without violating the bounds from other observables. In particular, the two leading scenarios~\cite{Alguero:2023jeh}, after the updated $R_K$ and $R_{K^*}$ measurements, are $(\mathcal{C}_{9\mu}^{\rm V}=-\mathcal{C}_{10\mu}^{\rm V},\mathcal{C}_9^{\rm U})$ with pulls of $5.8\sigma$ and $6.3\sigma$ (if $R(D^{(*)})$ is included), respectively. This suggests NP with a size of $\sim 20\%$ w.r.t.~the SM.

Since at least a dominant LFU NP effect is needed, direct NP contributions of LQs or new scalars and fermions via box diagrams would require a quite intricate tuning. This leaves $Z^\prime$ bosons and off-shell photon penguins induced via charm or tau loops as the most straightforward explanations, albeit the constraints from $B_s-\bar B_s$ mixing are rather stringent.

The $b\to c\tau\nu$ transitions, as tree-level mediated charged current process, have sizable decay rates and the ratio $R(D)$ and $R(D^*)$ point towards the violation of LFU. While it is possible to explain these tensions with scalar currents, at the same time slightly improving the polarization observables, the best fit is achieved via a NP contribution of $\sim 10\%$ to the SM operator with left-handed quarks and leptons, resulting in a significance of $\sim 4\sigma$. Such an operator can be generated in LQs models, preferably $U_1$ and $S_1+S_3$.

The solution of the $R(D)$ and $R(D^*)$ anomalies via a left-handed vector current also offers the most straightforward possibility of a combined explanation. In the particularly promising scenario $(\mathcal{C}_{9\mu}^{\rm V}=-\mathcal{C}_{10\mu}^{\rm V},\mathcal{C}_9^{\rm U})$, $\mathcal{C}_9^{\rm U}$ originates from a tau-loop while $\mathcal{C}_{9\mu}^{\rm V}=-\mathcal{C}_{10\mu}^{\rm V}$ comes from a direct tree-level effect. This can be naturally obtained with a $U_1$ LQ respecting a $U(2)$ flavour symmetry~\cite{Barbieri:2015yvd}. In this setup, one predicts the bin [1.1,6]~GeV bins~\cite{Alguero:2023jeh}:
\begin{eqnarray}
R_K=0.909\pm 0.060 \quad
R_{K^*}=0.914 \pm 0.053.\, \,
\label{predRk}
\end{eqnarray}
Furthermore, this scenario predicts measurable rates for $b\to s\tau^+\tau^-$ processes, liked to the size of $R(D^{(*)})$ and $\mathcal{C}_9^{\rm U}$. Finally, an explanation of $R(D^{(*)})$ via the $S_1$ LQ could also be linked \cite{Lizana:2023kei} to the emerging non-leptonic anomalies in $B$ decays.

\section*{Nota Added}
{After completion of this report, LHCb announced at the CKM2023 conference in Santiago de Compostela the result of a first unbinned analysis of $B\to K^* \mu^+\mu^-$  including a data-driven determination of charm loops effects~\cite{LHCbtalk}. The main outcome is that despite the extra degrees of freedom contained in the $c{\bar c}$ parameters, a NP contribution in ${\cal C}_9$ of the same size as the one found in~\cite{Alguero:2023jeh}, where charm-loop is taken from theory,  is needed to explain $P_5^\prime$. It is important to emphasize that this analysis includes only $B\to K^* \mu^+\mu^-$. Therefore it will be of utmost importance to see the global significance of a complete unbinned analysis adding electronic modes and all other channels like $B_s \to\phi\mu^+\mu^-$ and $B\to K\mu\mu$ with their corresponding charm parameterizations.
}

\section*{Acknowledgments}
A.C.~thanks Syuhei Iguro for useful discussions. A.C.~gratefully acknowledges the support by the Swiss National Science Foundation under Project No.\ PP00P21\_76884. JM acknowledges financial support from the Spanish Ministry of Science, Innovation and Universities (PID2020-112965GB-I00/AEI/ 10.13039/501100011033) and by ICREA under the ICREA Academia programme. The work of B.C. is supported by the Margarita Salas postdoctoral
program funded by the European Union-NextGenerationEU. 

\section*{Data Availability Statement}
Data Availability Statement: No Data associated in the manuscript.

\bibliography{sn-bibliography.bib}

\end{document}